\documentclass[twocolumn,aps,pre,preprintnumbers,showkeys,floatfix]{revtex4-2}

\usepackage[utf8]{inputenc}
\usepackage[colorlinks=true, allcolors=blue]{hyperref}
\usepackage{amsfonts}
\usepackage{amsmath}
\usepackage{amsthm}
\usepackage{dsfont}
\usepackage{enumitem}
\usepackage{tikz}
\usepackage{caption}
\usepackage{subcaption}

\usepackage{anyfontsize}
\usepackage{array}
\usepackage{natbib}

\newcommand{\avg}[1]{\langle{#1}\rangle}
\newcommand{\lavg}[1]{\left\langle{#1}\right\rangle}

\newcommand{\var}[1]{\mathrm{var}({#1})}

\newcommand{\JS}{J_S}
\newcommand{\J}{J}
\newcommand{\JRev}{J^{\text{rev}}}
\newcommand{\JIrr}{J^{\text{irr}}}
\newcommand{\JFull}{J^{\text{full}}}
\newcommand{\dbt}{d\boldsymbol{B}_{t}}
\newcommand{\boldx}{\boldsymbol{x}}
\newcommand{\boldp}{\boldsymbol{p}}

\newcommand{\At}{A_t}
\newcommand{\A}{A}

\newcommand{\up}{u_p}
\newcommand{\boldup}{\boldsymbol{u}_{p}}

\newcommand{\uRev}{\up^{\text{rev}}}
\newcommand{\uIrr}{\up^{\text{irr}}}
\newcommand{\bolduIrr}{\boldup^{\text{irr}}}
\newcommand{\bolduRev}{\boldup^{\text{rev}}}

\begin{document}

\title{Entropy Production by Underdamped Langevin Dynamics}
\author{Jinghao Lyu}
\email{jolyu@ucdavis.edu}
\affiliation{Complexity Sciences Center and Physics and Astronomy Department,
University of California at Davis, One Shields Avenue, Davis, CA 95616}

\author{Kyle J. Ray}
\email{kjray@ucdavis.edu}
\affiliation{Complexity Sciences Center and Physics and Astronomy Department,
University of California at Davis, One Shields Avenue, Davis, CA 95616}

\author{James P. Crutchfield}
\email{chaos@ucdavis.edu}
\affiliation{Complexity Sciences Center and Physics and Astronomy Department,
University of California at Davis, One Shields Avenue, Davis, CA 95616}

\date{\today}

\begin{abstract}
Entropy production (EP) is a central quantity in nonequilibrium physics as it monitors energy dissipation, irreversibility, and free energy differences during thermodynamic transformations. Estimating EP, however, is challenging both theoretically and experimentally due to limited access to the system dynamics. For overdamped Langevin dynamics and Markov jump processes it was recently proposed that, from thermodynamic uncertainty relations (TUR), short-time cumulant currents can be used to estimate EP without knowledge of the dynamics. Yet, estimation of EP in underdamped Langevin systems remains an active challenge. To address this, we derive a modified TUR that relates the statistics of two specific novel currents---one cumulant current and one stochastic current---to a system's EP. These two distinct but related currents are used to constrain EP in the modified TUR. One highlight is that there always exists a family of currents such that the uncertainty relations saturate, even for long-time averages and in nonsteady-state scenarios. Another is that our method only requires limited knowledge of the dynamics---specifically, the damping-coefficient to mass ratio and the diffusion constant. This uncertainty relation allows estimating EP for both overdamped and underdamped Langevin dynamics. We validate the method numerically, through applications to several underdamped systems, to underscore the flexibility in obtaining EP in nonequilibrium Langevin systems.


\end{abstract}


\preprint{arxiv.org:2405.12305}

\maketitle

\section{Introduction}

Over the past decade stochastic thermodynamics led to significant advances in our understanding of nonequilibrium systems. One highlight is the \emph{thermodynamic uncertainty relation} (TUR) that gives the trade-off between \emph{entropy production} (EP) $\Sigma$ (or \emph{dissipation}) and the resolution (or \emph{accuracy}) of a current $J$:
\begin{align}
    \var{J} \cdot \frac{\Sigma}{2} \geq \avg{J}^2~,
\end{align}
where $\var{J}$ and $\avg{J}$ are the variance and mean of $J$, respectively. 
After its first discovery in the nonequilibrium steady state (NESS) regime \cite{barato2015thermodynamic,gingrich2016dissipation,horowitz2017proof, horowitz2020thermodynamic}, the TUR was extended to non-NESS discrete Markovian systems \cite{liu2020thermodynamic}, overdamped Langevin systems \cite{koyuk2020thermodynamic}, underdamped Langevin systems \cite{lee2021universal}, active particles \cite{cao2022effective}, processes with measurement and feedback control \cite{potts2019thermodynamic}, and more \cite{pietzonka2016universal, polettini2016tightening, proesmans2017discrete,dechant2018current,barato2018bounds,macieszczak2018unified,brandner2018thermodynamic,koyuk2019operationally, chun2019effect,van2019uncertaintydelayed,dechant2018multidimensional,hasegawa2019fluctuation,guarnieri2019thermodynamics,proesmans2019hysteretic,van2020uncertainty,timpanaro2019thermodynamic,hasegawa2020quantum,ray2023thermodynamic}. Generally speaking, TURs provide lower bounds on the EP via observable currents. An array of techniques are used to derive these TURs, including large deviation theory \cite{gingrich2016dissipation}, the Cr\'amer-Rao bound \cite{hasegawa2019uncertainty}, the characteristic function inequality \cite{dechant2020fluctuation,koyuk2020thermodynamic}, and the Cauchy-Schwartz inequality \cite{dieball2023direct}.

EP is a key physical property of a thermodynamic transformation---often taken as a proxy for energy dissipation. It describes how irreversible a transformation is and quantifies the minimal thermodynamic resources to drive it. Saturating a TUR inequality allows estimating the EP (rate) directly from observed data. This reduces to an optimization problem: find a current $J$ that maximizes $2\avg{J}^2/\var{J}$. That maximum is the EP. Recently, in the short-time duration limit, it was proposed that equality holds for several specific currents \cite{li2019quantifying,manikandan2020inferring, otsubo2020estimating, van2020entropy, otsubo2022estimating}.  Using this, EP can be estimated without detailed knowledge of the forces in overdamped Langevin systems or the transition rates in Markovian systems. A natural follow-on then is: Can we use currents to estimate EP in other stochastic systems? Here, we focus on solving this problem for underdamped Langevin dynamics.

Several modified TURs for underdamped systems exist already \cite{van2019uncertainty,lee2021universal,fischer2020free,di2020thermodynamic,van2020thermodynamic}, but none are well suited to empirically estimating EP. The main difference in the underdamped case, compared to overdamped, is the appearance of reversible currents in the dynamics. And, these result in modified TURs with additional terms; for example, the total activity or the Fisher information. Some TURs also require settings with perfect control and knowledge of all protocol parameters. Such requirements are typically impossible to meet in experiments. Furthermore, the physical meaning and tightness of these modified TURs for underdamped systems are not clear, even in the short-time regime. 

The following directly addresses EP estimation in the underdamped regime. We introduce two different types of current---irreversible currents and stochastic currents---and derive a TUR that relates their moments to EP. We show that this TUR (i) applies to all Langevin (overdamped and underdamped) systems, (ii) works under arbitrary control protocols, (iii) is always saturable, and (iv) is not limited to short time or NESS inference. This TUR can be used to readily estimate EP from experimental measurements provided only that the system's damping-coefficient to mass ratio and the diffusion constant are known.


\section{Dynamics}

We consider a $n$-dimensional underdamped motion at the temperature $T$. The Langevin equations are:
\begin{align}
\label{eq:UD_langevin}
    d\boldsymbol{x} = \frac{\boldsymbol{p}}{m} dt ~,~ d\boldsymbol{p} = -\gamma \frac{\boldsymbol{p}}{m} dt + \boldsymbol{F}(\boldsymbol{x},t) dt + \dbt
    ~,
\end{align}
where $\boldsymbol{x}$, $\boldsymbol{p}$, $\boldsymbol{F}$, $\gamma$, and $m$ are the position, momentum, external force, damping coefficient, and particle mass, respectively. $\dbt$ are $n$-dimension independent infinitesimal Brownian motions---Wiener processes with variance $2Ddt$ 
where $D=\gamma k_{\mathrm{B}} T$ is the duffision constant. Throughout, we assume the Boltzmann constant $k_{\mathrm{B}}=1$.

The corresponding evolution of the probability distribution $f(\boldx,\boldp,t)$ from $t=0$ to $\tau$ is described by a continuity constraint called Kramer's equation:
\begin{align}
\label{eq:Kramers}
    \partial_{t}f(\boldsymbol{x},\boldsymbol{p},t)=-   \Big( \frac{\boldsymbol{p}}{m} \cdot\partial_{\boldsymbol{x}}f(\boldsymbol{x},\boldsymbol{p},t) +\partial_{\boldsymbol{p}}\left({\boldup}f(\boldsymbol{x},\boldsymbol{p},t)\right) \Big) ~,
\end{align}
where $\boldup= -\gamma \frac{\boldsymbol{p}}{m}+\boldsymbol{F}(\boldsymbol{x},t)-\gamma T \partial_{\boldsymbol{p}} \log f(\boldsymbol{x},\boldsymbol{p},t)$ is the probability velocity along $\boldsymbol{p}$. We decompose ${\boldup}$ into ${\bolduRev}$ and ${\bolduIrr}$ based on how they behave under the time reverse transformation $\boldsymbol{x} \to \boldsymbol{x}, \boldsymbol{p}\to -\boldsymbol{p}$: ${\bolduRev} \to {\bolduRev}$ and ${\bolduIrr} \to -{\bolduIrr}$. 
The following assumes ${\bolduRev}= \boldsymbol{F}(\boldsymbol{x},t)$ and ${\bolduIrr} = -\gamma \frac{\boldsymbol{p}}{m} - \gamma T \partial_{\boldsymbol{p}} \log f(\boldsymbol{x},\boldsymbol{p},t) $. 

We take the average EP of such a system to be the sum of the mean heat flux into the thermal environment and the change in Shannon entropy over the system's probability distribution function. 
The average EP can be written as:
\begin{align}
\label{eq:UD_Ent}
    \Sigma = \frac{1}{\gamma T} \int_{0}^{\tau} dt\int d\boldx d\boldp~ {\bolduIrr(\boldx,\boldp,t)^2}{f(\boldx,\boldp,t)}~.
\end{align}
 (See Appendix \ref{app:EP} for details.)
As we can see, the total EP is the sum of the sub-EP in each direction and generalizations to higher dimensions are straightforward given a method for estimating $\Sigma$ along any given direction. For clarity, we discuss only the sub-EP from here on out and assume a one-dimensional system.


\section{Currents and modified TURs}

While our primary focus is a method for estimating EP in underdamped systems, the underlying method applies to overdamped systems as well. We first consider currents and TURs in overdamped dynamics and then extend the method to underdamped systems.

\subsection{Overdamped dynamics}

 Analogous to Eq. \eqref{eq:UD_Ent}, in a one-dimensional overdamped system described by a probability distribution $f(x,t)$, the mean EP can be expressed as:
\begin{align}
    \Sigma = \frac{1}{D} \int_{0}^{\tau} dt \int dx f(x,t) u^2(x,t)~.
\end{align}
where $u(x,t)$ is the probability current velocity. A \emph{cumulant current} $\J(w)$ is the integral of a local weight function $w=w(x,t)$ over a particular trajectory $\Gamma$:
\begin{align}
    \J(w)&=\int_{\Gamma} w(x,t) \circ dx~,
\end{align}
where the $\circ$ represents a Stratonovich product. The average of this current over all trajectories is \cite{seifert2012stochastic}:
\begin{align}
    \avg{\J(w)} = \int_{0}^{\tau} dt \int dx f(x,t) u(x,t) w(x,t)~.
\end{align}
From this point forward, $\avg{\cdots}$ denotes averaging a trajectorywise quantity over all possible trajectories.

Inspired by Ref. \cite{dieball2023direct}, we define a conjugate \emph{stochastic current} $\JS$ with the same weight $w(x,t)$ as the physical current $\J$:
\begin{align}
    \JS(w)&=\int_{\Gamma} w(x,t) \cdot  dB_{t}~,
\end{align}
where  $\cdot$ denotes an Ito product. 
The stochastic current's statistical moments are imbued with special properties. First, 
$\JS$ has zero mean and so $\avg{\JS^2}=\mathrm{var}(\JS)$. Second, the covariance of the product of two stochastic currents is related to a particular cumulant current:
\begin{align}
\label{eq:Covariance}
    \lavg{\JS(w)\JS(w')} &= \lavg{\int_{\Gamma \times \Gamma} dB_{t_{1}}dB_{t_{2}}w(x_{t_{1}},t_{1})w'(x_{t_{2}},t_{2})} \nonumber \\
    &=2D \int_{0}^{\tau} dt \int dx ~f(x,t) w(x,t) w'(x,t)  \nonumber \\
    &=2D \avg{J(ww'/u)}~.
\end{align}
The relation above motivates a specific stochastic current, also defined in Ref. \cite{dieball2023direct}:
\begin{align*}
\At \equiv \frac{1}{{2D}} \JS(u)
  ~,
\end{align*}
which can be seen to satisfy both $\avg{\At^2} = \avg{J(u)} =\Sigma / 2$ and $\avg{\At \JS(w)} =\avg{J(w)} $  using the property above.

Now, suppressing the dependence on the specific $w$ and applying the Cauchy-Schwartz inequality, yields:
\begin{align}
    \avg{A_{t}^2} \avg{\JS^2} \geq \avg{\At \JS}^2 ~.
\end{align}
Plugging in the relations for $\At$ gives a TUR valid for any choice of $w(x,t)$:
\begin{align}
\label{eq:LTUR}
     \frac{\Sigma}{2} \geq \frac{\avg{\J}^2}{\mathrm{var}(\JS)}~.
\end{align}

 This inequality can be saturated if and only if $w(x,t)=c u(x,t)$, where $c$ is a constant. Also, note that there are two different types of current in our uncertainty relation. While $\JS$ is not a physically observable current, its variance can be readily calculated by using Eq. \eqref{eq:Covariance}:
\begin{align}
\label{eq:OD_average_stoch}
\var{\JS(w)} &= 2D \avg{J(w^2/u)} \nonumber \\
   & = 2D\lavg{\int_\Gamma w(x,t)^2 dt} ~.
\end{align}
This form shows that the quantity is as readily estimated from experimental trajectories as the cumulant current average. It also shows clearly that $\var{\JS}$ converges to $\var{J}$ as $\tau \to 0$, reducing Eq. \eqref{eq:LTUR} to previous results on short-time duration inference \cite{li2019quantifying,manikandan2020inferring, otsubo2020estimating, van2020entropy, otsubo2022estimating}.

\subsection{Underdamped}

The same approach yields a TUR for the underdamped regime. Naively, we simply promote our trajectories, integrals, and weight functions to be defined over the $(x,p)$ phase space: $w = w(x,p,t)$ and cumulant currents $\int w(x,p,t) \circ dp$. We maintain the same definition of $\JS=\int_{\Gamma} w(x,p,t) \cdot dB_{t}$ and, by looking at Eq. \eqref{eq:UD_Ent}, we define $\A = J_s(\uIrr)/2D$ to recover $\avg{\A^2} = \Sigma / 2$.

Continuing as before, we investigate the average of the product of $A$ and $\JS$, finding that $\avg{\A \JS(w)} = \avg{ w \uIrr}$. However, this is not the average of the naively promoted cumulant current $\int w(x,p,t)\circ dp$. Thus, to display the necessary average behavior, we require a current such that $\langle \J(w) \rangle = \avg{ w \uIrr}$---one that only accounts for the irreversible part of the probability current, rather than the full probability current. The naive extension of our overdamped current fails, but the following definition suffices:
\begin{align}
\label{eq:UDJ}
    \J(w)&=\int_{\Gamma} w(x,p,t) \gamma dx -\frac{1}{2} \int_{\Gamma} dw \cdot dp~ 
\end{align}
(See App. \ref{app:J_irr} for more details). 
Again, using the Cauchy-Schwartz inequality:
\begin{align}
    \avg{A^2} \avg{\JS^2} \geq \avg{A\JS}^2~.
\end{align}
We recover Eq. \eqref{eq:LTUR} where the equality holds if and only if $w(x,p,t) \sim \uIrr(x,p,t)$. Notably, this TUR holds for completely arbitrary over- or underdamped Langevin dynamics (over any timescale for any time-dependent driving force and any initial probability). It is not limited by being valid only for certain regimes, initial conditions, timescales, or driving forces. 

\begin{figure*}[t]
    \begin{subfigure}{1\columnwidth}
    \centering
    \scalebox{0.51}{
    \includegraphics{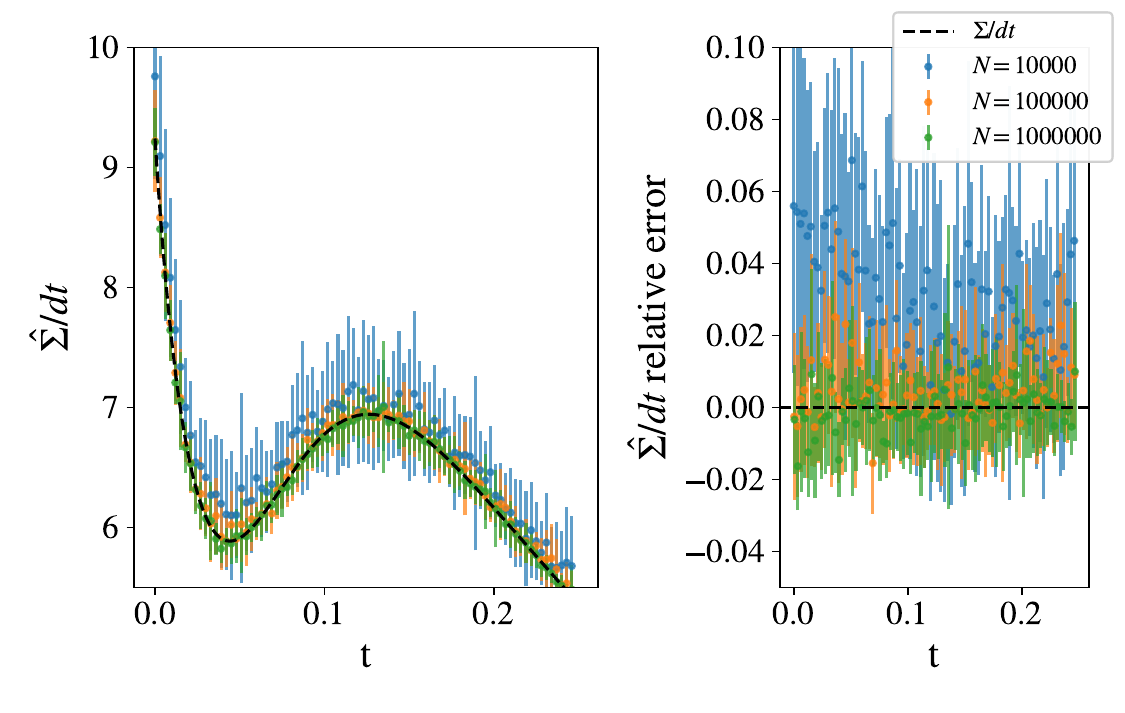}
    }
    \caption{}
    \label{fig:udfreediffusion_10by10}
    \end{subfigure}
    \begin{subfigure}{1\columnwidth}
    \centering
    \scalebox{0.5}{
    \includegraphics{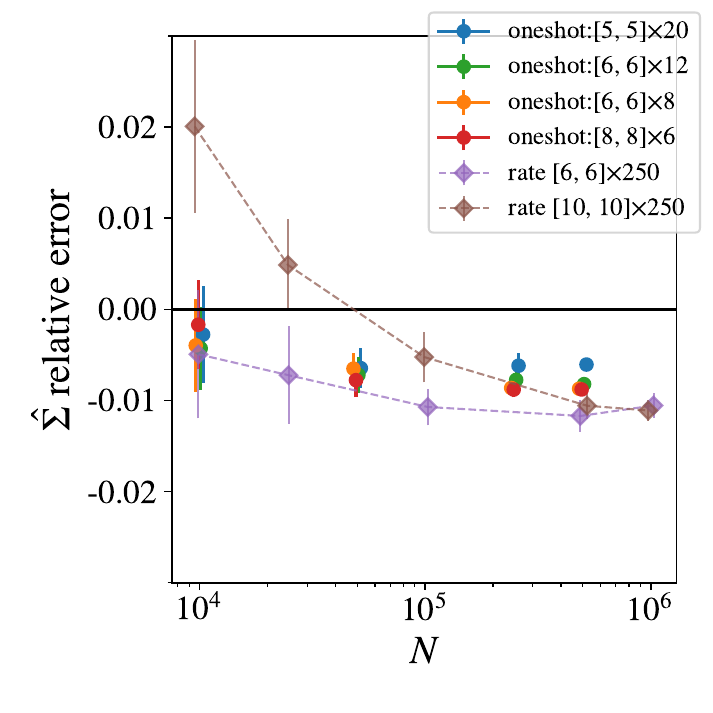}
    }
    \caption{}
    \label{fig:oneshot}
    \end{subfigure}
\caption{(a) EP rate in free diffusion of an underdamped particle. All dynamical
    parameters are set to unity. The initial Gaussian distribution parameters are $(p_{0}, \sigma_{p_{0}}, \sigma_{x_{0}})=(0.2, 0.3, 0.025)$. We select $\approx 80$ time steps to infer the EP rate. For each time step, $10\times10$ Gaussian kernels in the phase space $(x,p)$ are used. We simulated $10^4$ (blue), $10^5$ (orange), and $10^6$ (green) trajectories with time length $\tau=0.25$ and $\Delta t=10^{-3}$. The black dashed line is the theoretical EP rate as a function time $t$. The rightmost figure shows the relative error based on $20$ independent trajectory ensembles. We see that the model is mostly converged by the time it gets $10^5$ data points. (b) Relative error for total EP estimation over the entire trajectory. Both one-shot (circles) and rate-based (diamonds) estimates are plotted, using various numbers of estimators in the $x$, $p$, and $t$ directions. Note that the rate-based estimations require more data to converge, reaching errors of about a percent. The notation in the legend refers to $[n_{x_i},n_{p_i}]\times n_{t_i}$ basis functions used. In all cases, trajectories with $250$ time steps were used.
    }
\end{figure*}

    %

\section{Estimating EP}

In experiments, we observe discretized trajectories of Langevin dynamics: $\Gamma_{\text{dis}} = \{(x_i,p_i/m,t_i)\}_{i=0}^{N}$, where $x_i$ and $p_i$ are the position and momentum at time $t_{i}= i \cdot \tau/N$.
Suppose we observe $N_{\Gamma}$ trajectories and pick a weight function $w(x,p,t)$. The average $\avg{\J}$ and $\var{\JS}$ can be estimated by averaging $\J_{\Gamma}$ of each trajectory. 
In underdamped systems:
\begin{align}
     & \avg{J} = \\
     &\lavg{\sum_{i=0}^{N-1}\Big( \gamma w_i (x_{i+1}-x_i)
    -\frac{1}{2}(w_{i+1}-w_i)(p_{i+1}-p_i)\Big)} \nonumber \\
    & \var{J_{S}}  =2D\lavg{\sum_{i=0}^{N-1}w_{i}^2 dt}
    ~,
\end{align} 
where $w_i$ is the weight function evaluated at $(x_i,p_i,t_i)$.

In this way, estimating EP becomes an optimization problem:
\begin{align}
    \widehat{\Sigma} = \max_{w} \frac{2\avg{\J}^2}{\mathrm{var}(\JS)}~.
\end{align}
Compared to the method used in Refs. \cite{van2020entropy,dechant2019thermodynamic,otsubo2020estimating,otsubo2022estimating}, our bound is always saturable for any time dependent drive over any time scale. However, we also need additional knowledge: the damping coefficient, the mass and the diffusion constant. Since we can rescale $w(x,p,t)$ freely, only the damping-coefficient to mass ratio $\gamma/m$ and diffusion constant are needed. 

Finding the optimal expression of $w(x,p,t)$ for an arbitrarily time-varying system is challenging in general. A good model class for such a weight $w$ could contain millions of parameters. 
To demonstrate our results, though, we follow Ref. \cite{van2020entropy} and use a set of Gaussian basis functions $\{\phi_{i}\}_{i}$ to approximate the weight.

There are two ways to expand general time-dependent weights $w(x,p,t)$. We can let either coefficients or basis functions have time dependence:
\begin{align}
    w(x,p,t) \sim \sum_{i}
        \Bigl\{
    \begin{array}{l}
       c_{i} \cdot \phi_{i}(x,p,t) ~ (t\text{ in basis functions})\\
       c_{i}(t) \cdot \phi_{i}(x,p) ~(t\text{ in coefficients})~.
       \end{array}  
\end{align}
We refer to estimations based on these two expansions as \emph{rate-based estimations} and \emph{one-shot estimations}, respectively. The meaning of these names will become clear.

By picking basis functions to approximate the weight, the optimal weight $w_{\mathrm{opt}}$ for a given sample of trajectories can be written in a closed form. To avoid overfitting to small data sets, we also include $L\text{-}2$ norm regularization. (Details for the exact regularized solution can be found in Appendix \ref{appendix:optimalclosedresult}.) In all cases, we use equal-width multivariate Gaussian basis functions, with centers placed uniformly across the data domain. We make this naive choice so as to not allow our method to be particularly adapted to the examples below---an attempt to simulate working with data for which the underlying dynamics and distribution are truly unknown.

\subsection{Rate-based estimation}

If coefficients depend on time $t$, we find the optimal coefficients for each discretized time $t_{i}$ independently. Appendix \ref{appendix:optimalclosedresult} shows this estimation is equivalent to using the short trajectories from time $t_{i}$ and $t_{i+1}$ to infer the EP during a short period of time and summing all EPs together. With this, we can divide the long-time trajectories into subtrajectories of length $\Delta t$ to infer the EP. For each subtrajectory from $t_{i}$ to $t_{i+1}$, dividing the EP with time separation $\Delta t$ gives the EP rate at $t_{i}$. This is why we call it rate-based estimation. This expansion is similar to previous approaches that used short-duration currents to infer time-dependent EP rates \cite{otsubo2022estimating}. 

\subsection{One-shot estimation}

Our TUR is saturable for any time-length trajectory. 
The previous expansion does not show this advantage of our TUR, as we measure the EP rate with short-duration currents essentially. This raises a question: Can we estimate the total EP directly instead of EP rates? This suggests using basis functions that span the whole spacetime, by letting basis functions depend on time $t$. Compared to estimating EP rate for each time step, this significantly reduces the number of basis functions needed to estimate the total EP as we do not use different basis functions for each time step. That being said, the number of calculations to arrive at the final $\Sigma$ is not necessarily smaller for the method at hand, because the Gaussian time kernels have infinite range. More intelligent kernels or a different optimization algorithm would likely improve the one-shot method.

\section{Numerical Experiments}

The first test of EP estimation concerns unconstrained diffusion. The second explores a particle constrained to move on a ring.

\subsection{Free diffusion}

Our first example is a simple but important case: vanishing external force $\boldsymbol{F}=0$ in one dimension.
The initial distribution is set to be Gaussian in both $x$ and $p$.
The EP rates' trend as a function of time is different based on the parameters of the initial Gaussian distribution. In Fig. \ref{fig:udfreediffusion_10by10}, we calculate EP for initial conditions that generate a nonmonotone EP rate. 

First, we use rate-based estimation to infer the EP rate.
At each time step, we choose $100$ Gaussian kernels to cover the phase space uniformly.
As seen in Fig. \ref{fig:udfreediffusion_10by10}, our method produces good estimates of the EP rates at different time steps. 

We can also estimate the total EP $\Sigma$ over the interval using either the rate-based or the one-shot method. Figure \ref{fig:oneshot} compares the methods. Notably, the one-shot method appears superior under the constraint of limited data; likely as it acts as an additional regularizer that prevents over-fitting to the variations between each successive time step. However, it does not appear to provide much advantage asymptotically.

\subsection{Particle on a ring}

Next, we explore a simulation of the NESS state of an underdamped particle on a ring under a time-independent drive (system details in appendix \ref{app:ParticleOnRing}). 
While lacking a closed form NESS distribution, we can use the explicit force expression and simulated trajectories to estimate average EP directly, establishing a ground truth. In the TUR estimations, we assume no knowledge of the forcing function, only having access to trajectories.

\begin{figure}
    \centering
    \includegraphics[width=\columnwidth]{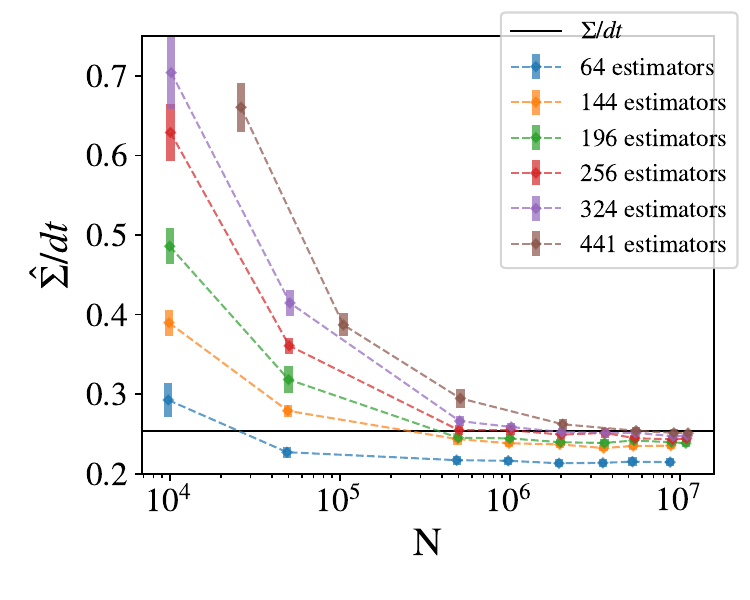}
     \caption{Cyclic NESS EP rate estimation: We estimate the EP with different numbers of estimators and trajectories. With low number of estimators (64, 144, 196), the results converge fast with less than $\sim10^6$ trajectories. When we increase the number of estimators, the convergent values are closer to the real EP. However, we need more trajectories for the estimated EP rate to converge.}
    \label{fig:cyclic_NESS18N126}
\end{figure}

    %
Figure \ref{fig:cyclic_NESS18N126} illustrates the results for this cyclic underdamped particle. For each trial, we extract data for step $t_{i}$ and step $t_{i+1}$ for all trajectories. We compute the optimal weight $w_{opt}$ for this ensemble of two successive steps---and use the EP to get the EP rate. 
The EP rate is estimated with different numbers of estimators for different numbers of paths from $10^4$ to $10^7$ . In general, too few estimators results in an asymptotic ``underfitting'' of the data as the model does not have sufficient flexibility to capture the true $w(x,p,t)$. More estimators yield more exact estimation after convergence, but this is also prone to overfitting until a large numbers of realizations are provided. Figure \ref{fig:cyclic_NESS18N126} shows that the standard error of the estimate decreases as the number of paths increases for any fixed number of estimators. With sufficient estimators, the estimated EP rate converges to the EP rate calculated directly. 

\section{Conclusion}

We generalized previous Langevin TURs that used Stratonovich product-based cumulant currents with weights to stochastic currents and irreversible currents. We derived modified TURs that relate statistics of these two currents to the EP. The modified TURs hold in both over- and underdamped systems and for arbitrary initial distributions, protocols, and timescales. Furthermore, the TURs can always be saturated. 

We then used our TURs to study EP numerically for several underdamped Langevin examples by approximating the weight with Gaussian kernels. We studied two different methods to estimate EP---rate-based and one-shot estimation. The first only uses short-duration trajectories and the second uses entire trajectories. As a result, rate-based estimation yields the EP rate while one-shot estimation leads to a process' total EP.

This generalization opens a new window to constructing currents that probe different quantities of interest in Langevin dynamics. 
Additionally, our method provides a framework to estimate the EP in modified Langevin dynamics that include, for example, active Brownian motion, position-dependent damping coefficient, and the like.

Our treatment shares several aspects with the method used in Ref. \cite{lee2023multidimensional}, though here we investigated the EP from trajectories rather than distributions. We focused exclusively on underdamped examples---for which EP estimation is much more challenging. Our unregularized method to find the optimal weight is based on the algorithm in Ref. \cite{van2020entropy}. 

Finding a better optimization algorithm will be an important step in future applications of this TUR in higher dimensional systems. Using a stochastic method and smarter basis function choices are all potential avenues for improvement. Additionally, machine learning methods have already been used to avoid intensive computations in estimating the EP \cite{otsubo2020estimating,kim2020learning,kwon2024alpha}. The successes reported here suggest that machine learning methods may assist in estimating the EP of underdamped dynamics as well.


\section*{Acknowledgments}

The authors thank Alec Boyd for helpful discussions. KJR and JPC thank the Telluride Science Research Center for hospitality during visits and the participants of the Information Engines Workshops there. This material is based upon work supported by, or in part by, U.S. Army Research Laboratory, U.S. Army Research Office grant W911NF-21-1-0048.

\bibliography{ref}

\clearpage

\appendix

\section{EP in underdamped dynamics}
\label{app:EP}

This following dvelops several main results in underdamped Langevin dynamics. Consider an $m-$dimensional particle with initial distribution $f(\boldx,\boldp,0)$ at $t=0$ and governed by the Langevin equations: 
\begin{align}
    d\boldsymbol{x} & = \frac{\boldsymbol{p}}{m} dt \nonumber \\
    d\boldsymbol{p} & = -\gamma \frac{\boldsymbol{p}}{m} dt + \boldsymbol{F}(\boldsymbol{x},t) dt + \dbt~.
\end{align}

The heat in Langevin dynamics can be defined for each short trajectory:
\begin{align}\label{eqn:appheatdirectdef}
    -dQ &=  \left(-\gamma\frac{\boldp}{m}dt + d\boldsymbol{B}_{t}\right)\circ \frac{d\boldx}{dt} \nonumber \\
    & =-\gamma\frac{\boldp}{m} \circ d\boldx + \frac{\boldp}{m}\circ d\boldsymbol{B}_{t}\nonumber\\
    &=\frac{\boldp}{m} \circ d\boldp- \boldsymbol{F}(\boldx,t)\cdot{\frac{\boldp}{m}dt}
    ~,
\end{align}
where the first term tracks the work done by the damping and the second monitors the distribution from the random force imposed the thermal bath. The average heat is given by averaging among all trajectories:
\begin{align}
    -d\avg{Q} &= dt \int d\boldx d\boldp(\frac{\boldp}{m}\cdot \boldup-\boldsymbol{F}\cdot\frac{\boldp}{m})f(x,p,t)\nonumber \\
    &=dt\int d\boldx d\boldp \frac{\boldp}{m}\cdot \bolduIrr f(x,p,t)
    ~.
\end{align}
(Recall that $\boldup= -\gamma \frac{\boldsymbol{p}}{m}+\boldsymbol{F}(\boldsymbol{x},t)-\gamma T \partial_{\boldsymbol{p}} \log f(\boldsymbol{x},\boldsymbol{p},t) = \bolduIrr + \boldsymbol{F}$).)

The average EP is still defined as $d S + d\avg{Q}/T$, where $d S$ is the Shannon entropy change in the system:
\begin{align}\label{eqn:appendixdS}
    dS &= -dt \int d\boldx d\boldp ~\partial_{t}f(\boldx,\boldp,t)\log f(\boldx,\boldp,t) \nonumber \\
    &=  dt \int d\boldx d\boldp \Big( \frac{\boldsymbol{p}}{m} \cdot\partial_{\boldsymbol{x}}f(\boldsymbol{x},\boldsymbol{p},t) +\partial_{\boldsymbol{p}}\left({\boldup}f(\boldsymbol{x},\boldsymbol{p},t)\right) \Big)\nonumber\\
    &\log f(\boldx,\boldp,t)\nonumber\\
    &=  dt \int d\boldx d\boldp \partial_{\boldp}( \bolduIrr f(\boldsymbol{x},\boldsymbol{p},t)) \log f(\boldsymbol{x},\boldsymbol{p},t)\nonumber \\
    &= - dt \int d\boldx d\boldp  \bolduIrr f(\boldsymbol{x},\boldsymbol{p},t)\partial_{\boldp}\log f(\boldsymbol{x},\boldsymbol{p},t),
\end{align}
where from the second line to the third line we use that $\frac{\boldsymbol{p}}{m} \cdot\partial_{\boldsymbol{x}}f(\boldsymbol{x},\boldsymbol{p},t)\log f(\boldx,\boldp,t)$ and $\partial_{\boldsymbol{p}}\left(\boldsymbol{F}(\boldx,t)f(\boldsymbol{x},\boldsymbol{p},t)\right) \log f(\boldx,\boldp,t)$ do not contribute to the integral since they are surface integrals. Piecing everything together leads to:
\begin{align}
    \Sigma =& d\avg{Q}/T\nonumber+ dS\\
    =& dt\int d\boldx d \boldp~ \left(-\frac{p}{mT}\bolduIrr -  \bolduIrr \cdot \frac{1}{\gamma T}( -\bolduIrr - \frac{\gamma p}{m}) \right)f(\boldx,\boldp,t) \nonumber\\
    =& \frac{dt}{\gamma T}\int d\boldx d \boldp~ {(\bolduIrr)}^2 f(\boldx,\boldp,t) \nonumber\\
    =& \frac{dt}{\gamma T}\int d\boldx d \boldp~ \nonumber \\
    & \left(-\gamma \frac{\boldsymbol{p}}{m} - \gamma T \partial_{\boldsymbol{p}} \log f(\boldsymbol{x},\boldsymbol{p},t)\right)^2{f(\boldx,\boldp,t)}~.
\end{align}
As we expect, the average EP is always nonnegative.

There is an alternative way to use trajectory probabilities to define the EP for each trajectory. In underdamped Langevin dynamics, these two definitions lead to the same EP in the ensemble average. We first study the probability of finding the particle at $(\boldsymbol{x}+d\boldsymbol{x},\boldsymbol{p}+d\boldsymbol{p})$ at time $t+dt$ given that the particle is at $(\boldx,\boldp)$ at time $t$ in the original dynamics. From Langevin equation, we must have $d\boldx = \boldp/m dt$. The infinitesimal Brownian motion obeys the Gaussian distribution with 0 mean and $2D dt$ variance. In each direction $\alpha$, from $t$ to $t+dt$, the $\alpha$-th component of the infinitesimal Brownian motion takes the value $y_{\alpha}=dp_{\alpha}+\gamma \frac{p_{\alpha}}{m}  dt - F_{\alpha}(\boldx,t) dt$. This gives us the conditional distribution:
\begin{align}
    \mathrm{Pr}(\boldx&+d\boldx,\boldp+d\boldp, t+dt|\boldx,\boldp,t) \nonumber \\
    & = \prod_{\alpha=1}^{m} \frac{1}{\sqrt{2 \pi 2 D dt}} 
    \exp\left( -\frac{y_{\alpha}^2}{4 Ddt}\right)
    ~.
\end{align}

For each dynamics, the corresponding time-reversed dynamics is the same except that the initial distribution is $f(\boldx,-\boldp,\tau)$ and the force is 
$\boldsymbol{F}(\boldsymbol{x},\tau-t)$, where $\tau$ is the final time of the transformation.
For the infinitesimal trajectory $\Gamma_{dt}$: $(\boldx, \boldp, t) \to (\boldx+d\boldx,\boldp+d\boldp,t+dt)$, the conjugate time-reversed trajectory $\Gamma^{\dagger}_{dt}$ is $ (\boldx+d\boldx,-\boldp-d\boldp,t) \to (\boldx,-\boldp,t+dt)$ with time-reversed dynamics. We denote the probability in the time-reversed dynamics as $\mathrm{Pr}^{\dagger}(\cdot)$.
The EP assigned to each infinitesimal trajectory $\Gamma_{dt}$ is:
\begin{align}
    \Sigma_{\Gamma_{dt}} &= \log \frac{\mathrm{Pr}(\Gamma_{dt})}{\mathrm{Pr}^{\dagger}(\Gamma^{\dagger}_{dt})}\nonumber \\
    &=\log \frac{\mathrm{Pr}(\boldx+d\boldx,\boldp+d\boldp,t+dt|\boldx,\boldp,t)}{\mathrm{Pr}^{\dagger}(\boldx,-\boldp,t+dt|\boldx+d\boldx,-\boldp-d\boldp,t)}\nonumber \\
    &\qquad \times \frac{f(\boldx,\boldp,t)}{f^{\dagger}(\boldx+d\boldx,-\boldp-d \boldp,t)}~ \nonumber\\
    =& \log \frac{f(\boldx,\boldp,t)}{f(\boldx+d\boldx,\boldp+d\boldp,t+dt)} + \frac{dQ}{T}~,
\end{align}
where the infinitesimal heat is:
\begin{align}\label{eqn:appheatprobdef}
    dQ=\log \frac{\mathrm{Pr}(\boldx+d\boldx,\boldp+d\boldp,t+dt|\boldx,\boldp,t)}{\mathrm{Pr}^{\dagger}(\boldx,-\boldp,t+dt|\boldx+d\boldx,-\boldp-d\boldp,t)}~.
\end{align} 
We know that:
\begin{align*}
    \mathrm{Pr}(\boldx+d\boldx,\boldp+d\boldp,t+dt|\boldx,\boldp,t) & \sim \exp\left( -\frac{y_{\alpha}^2}{4 Ddt}\right)\\
    \text{and} \nonumber \\
    \mathrm{Pr}^{\dagger}(\boldx,-\boldp,t+dt|\boldx+d\boldx,-\boldp-d\boldp,t) & \sim \exp\left( -\frac{y_{\alpha}'^2}{4 Ddt}\right)
\end{align*}
where:
\begin{align*}
    y_{\alpha}&=dp_{\alpha}+\gamma \frac{p_{\alpha}}{m}  dt - F_{\alpha}(\boldx,t) dt \\
    & \text{and} \\
    y_{\alpha}'&=dp_{\alpha}-\gamma \frac{p_{\alpha}+dp_{\alpha}}{m}  dt - F_{\alpha}(\boldx+d\boldx,t+dt) dt~.
\end{align*}
And, the definition in Eq. \eqref{eqn:appheatprobdef} leads to the same heat defined in Eq. \eqref{eqn:appheatdirectdef} for infinitesimal trajectories. For a more comprehensive review on this topic, we refer readers to Refs. \cite{sekimoto2010stochastic, spinney2012entropy,peliti2021stochastic}.
\section{Average behavior of underdamped J}
\label{app:J_irr}

Definition Eq. \eqref{eq:UDJ} can be understood in two different ways. First, we can see by explicit averaging that it shows the proper average behavior. For the Ito part of the integral, we note that $dw \cdot dp \sim \partial_{p}w dp \cdot dp \sim 2D \partial_{p}w dt $, and write the average as:
\begin{align*}
    \avg{\J} &= \int dtdxdp~ w \frac{\gamma}{m} p f -  \int dtdxdp~ D \partial_{p}w f \\
    &= \int dtdxdp~ w \frac{\gamma}{m} p f + \int dtdxdp~ D w \frac{\partial_{p}f}{f} f\\
    &=\int dtdxdp (w~\uIrr) f
    ~,
\end{align*}
where we assume boundary terms vanish due to the distribution function being bounded appropriately. This last term is simply an expression for $\langle w \uIrr\rangle$, showing the current in a way that suits our purposes, which only reflects the irreversible current part.

Another perspective comes from attempting to use the full current $\JFull$; that includes contributions from $\uIrr$ and $\uRev$. We decompose $\JFull$ into two parts:
\begin{align}
    \JFull \equiv& \int_{\Gamma} w(x,p,t) \circ dp \nonumber \\
        =& \int_{\Gamma} - \gamma w(x,p,t) dx + w(x,p,t) \circ dB_{t} \nonumber \\
        &+ w(x,p,t) F(x,t)dt \nonumber\\
        =& \int_{\Gamma} - \gamma w(x,p,t) dx + D\partial_{p}w(x,p,t) dt \nonumber \\
        &+ w(x,p,t) \cdot dB_{t}+ w(x,p,t) F(x,t)dt \nonumber\\
        =&\JIrr+ \JRev~,
\end{align}
where:
\begin{align}
\JIrr = \int_{\Gamma} - \gamma w(x,p,t) dx & + D\partial_{p}w(x,p,t) dt \nonumber \\
   & + w(x,p,t) \cdot dB_{t}
\end{align}
and:
\begin{align}
\JRev=\int_{\Gamma}w(x,p,t) F(x,t)dt
~.
\end{align}
The average of $\J$ is equal to the average of $\JIrr$, since the last term in $\JRev$ vanishes upon averaging. Thus, our current $\J$ captures the irreversible part of the process and can be used to estimate the EP, since the EP only depends on $\JIrr$ and not $\JRev$.

\begin{figure*}[t]
    \begin{subfigure}{1\columnwidth}
        \centering
        \caption{}
        \scalebox{0.45}{
        \includegraphics{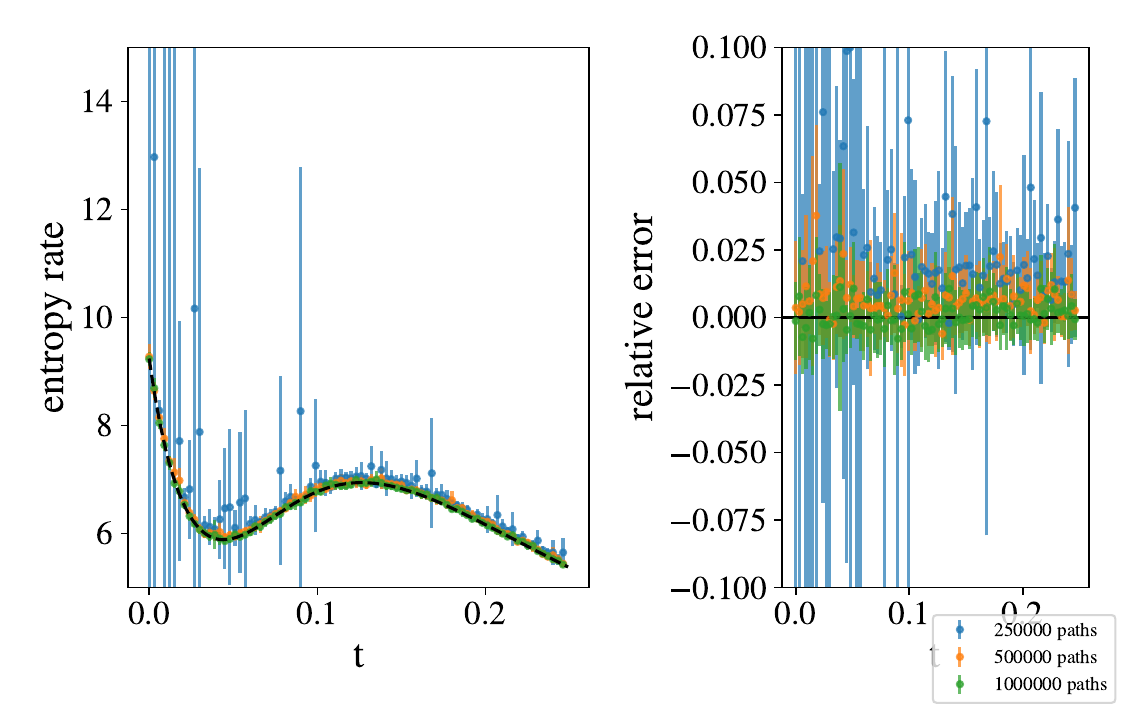}
        }
        \label{fig:rate_reg}
    \end{subfigure} 
    \begin{subfigure}{1\columnwidth}
        \centering
        \caption{}
        \scalebox{0.45}{
        \includegraphics{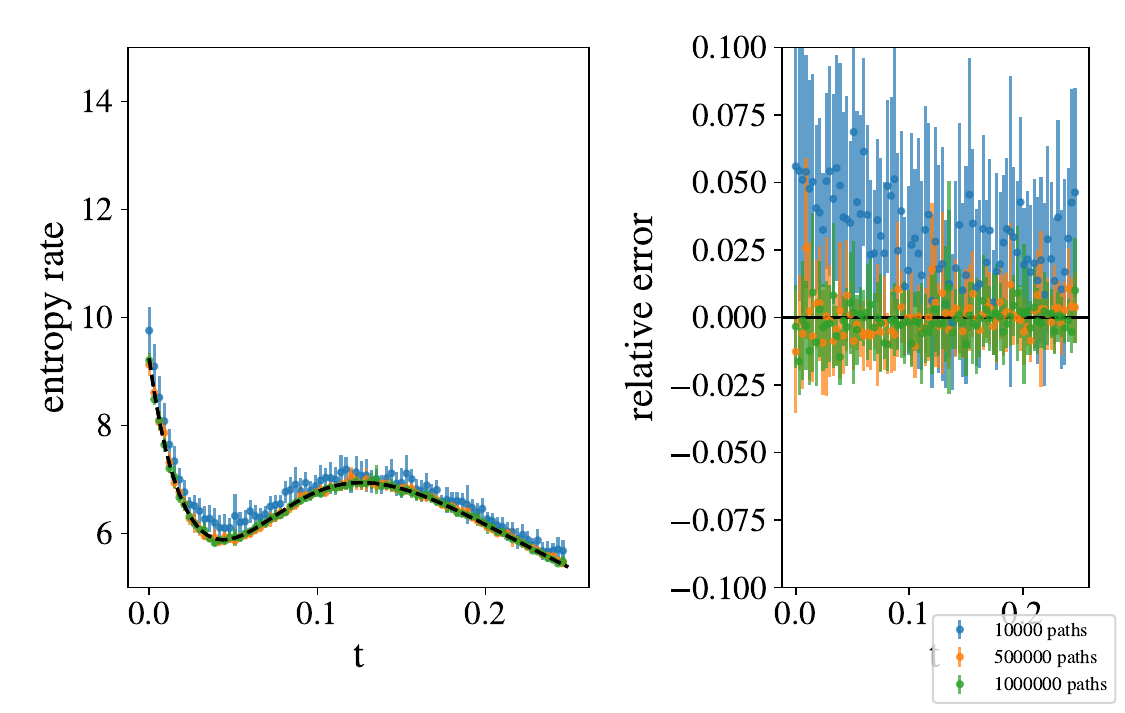}
        }
        \label{fig:rate_unreg}
    \end{subfigure}
    \caption{Results for the (a) unregularized and (b) regularized estimation. The blue data represents the behavior when given 250,000 (10,000) data points for the unregularized (regularized) models. Orange and green data are 500,000 and 1,000,000 data points, respectively, in both panels. Note that without the L2 norm regularization, the results are very messy even for 250,000 data points while the regularized data is already well behaved, if biased, for only 10,000 data points (compare blue to blue). The converged behavior is quite similar (compare green to green), but the regularized model reaches convergence must faster. }
    \label{rate_reg_unreg}
\end{figure*}

\section{Closed-form desired weights}
\label{appendix:optimalclosedresult}

This section introduces the closed form of the desired weight as originally derived from Ref. \cite{van2020entropy}. We start with $m-$dimensional NESS states where the weight $\boldsymbol{w}(\boldsymbol{x},\boldsymbol{p})$ does not depend on time $t$. The weight function $\boldsymbol{w}$ also has $m$ components. We use $k$ basis functions to approximate each component of the weight:
\begin{align*}
    w_{\alpha}(\boldsymbol{x},\boldsymbol{p}) = \sum_{i=1}^{k} c_{i\alpha}\phi_{i}(\boldsymbol{x},\boldsymbol{p})~,
\end{align*}
which follows the notation from Ref. \cite{van2020entropy} and where $\alpha$ goes from $1$ to $m$. The variance of the stochastic current with weight $w_{\alpha}(\boldsymbol{x},\boldsymbol{p})$ can be written as:
\begin{align*}
    \frac{1}{2D} \var{\JS} &= \lavg{\int_{\Gamma} w_\alpha^2 dt} \\
    &=\sum_{i,j} c_{i\alpha}c_{j\alpha} \lavg{\int_{\Gamma} \phi_{i}\phi_{j} dt} \\
    &=\sum_{i,j} c_{i\alpha}c_{j\alpha}\Xi_{ij}
    ~,
\end{align*}
where:
\begin{align*}
    \Xi_{ij}=\lavg{\int_{\Gamma} \phi_{i}\phi_{j} dt}~
\end{align*}
is a $k\times k$ symmetric matrix. The average of the irreversible current in $\alpha-$th direction is:
\begin{align*}
    \lavg{{\JIrr}_{\alpha}} &= \lavg{\int_{\Gamma} w \gamma dx_\alpha-\frac{1}{2} \int_{\Gamma} dw_\alpha \cdot dp_\alpha} \\
    &=\sum_{i} c_{i\alpha} \lavg{\int_{\Gamma} \phi_{i} \gamma dx -\frac{1}{2} \int_{\Gamma} d\phi_{i} \cdot dp_\alpha} \\
    &=\sum_{i} c_{i\alpha}\mu_{i\alpha}~,
\end{align*}
where:
\begin{align*}
    \mu_{i\alpha} = \lavg{\int_{\Gamma} \phi_{i} \gamma dx -\frac{1}{2} \int_{\Gamma} d\phi_{i} \cdot dp_\alpha}~.
\end{align*} 
The ratio $\mathcal{J}_{\alpha}$ we want to maximize with respect to
$\{c_{i}\}_{i}$ is:
\begin{align*}
    \mathcal{J}_{\alpha} = \frac{\sum_{i,j} c_{i}c_{j}\mu_{i\alpha}\mu_{j\alpha}}{\sum_{i,j} c_{i}c_{j}\Xi_{ij} }~.
\end{align*}
This can be done by directly asking the partial derivatives to vanish which leads to the maximum of $\mathcal{J}_\alpha$:
\begin{align}
    {\mathcal{J}_\alpha}_{\mathrm{max}} = \sum_{i,j} \mu_{i\alpha} \Xi^{-1}_{ij} \mu_{j\alpha}~.
\end{align}
We achieve the maximum when $c_{i\alpha}\sim \sum_{j}\Xi^{-1}_{ij} \mu_{j\alpha}$, where $\Xi^{-1}$ is the inverse of $\Xi$. The $\sim$ sign comes from $\mathcal{J}$ being invariant under rescaling $c$. For multidimensional system, we need to compute the $\Xi$ matrix once and $\mu$ for each direction. 

For nonNESS states, we first choose that our coefficients $c_{i\alpha}$ depend on time $t$ which we denote as $c_{i\alpha}(t_{n})$ at time step $t_{n}$:
\begin{align}
    w_{\alpha}(\boldsymbol{x},\boldsymbol{p},t) = \sum_{i}^{k} c_{i\alpha}(t)\phi_{i}(\boldsymbol{x},\boldsymbol{p})~.
\end{align}
The variance of current $\JS$ can be written as:
\begin{align}
    \var{{\JS}_\alpha}= \lavg{\int_{t_0}^{t_{1}} w_\alpha^2 dt}+\cdots+\lavg{\int_{t_{N-1}}^{t_{N}} w_\alpha^2 dt}~.
\end{align}
Discretizing this by taking the value of the starting point of each integral:
\begin{align}
    \var{{\JS}_\alpha}_{\text{start}}&= c_{i\alpha}(t_0)c_{j\alpha}(t_0)\lavg{\int_{t_0}^{t_{1}} \phi_{i}\phi_{j} dt} + \cdots \nonumber \\
    &+c_{i\alpha}(t_{N-1})c_{j\alpha}(t_{N-1})\lavg{\int_{t_{N-1}}^{t_{N}} \phi_{i}\phi_{j} dt}~.
\end{align}
Using ending point discretization leads us to:
\begin{align}
    \var{{\JS}_\alpha}_{\text{end}}&= c_{i\alpha}(t_1)c_{j\alpha}(t_1)\lavg{\int_{t_0}^{t_{1}} \phi_{i}\phi_{j} dt} + \cdots \nonumber \\
    &+c_{i\alpha}(t_{N})c_{j\alpha}(t_{N})\lavg{\int_{t_{N-1}}^{t_{N}} \phi_{i}\phi_{j} dt}~.
\end{align}
These two are equal to each other as $dt\to0$ since we integrate with time $dt$. To make the expression more symmetric, we average these two:
\begin{align}
    \var{{\JS}_{\alpha}} = \frac{1}{2}\var{{\JS}_{\alpha}}_{\text{start}} + \frac{1}{2}\var{{\JS}_{\alpha}}_{\text{end}}~.
\end{align}
It can be written as:
\begin{align}
    \var{\JS} &= \frac{1}{2} \sum_{i} c_{i\alpha}(t_0)c_{j\alpha}(t_0) \Xi(t_0)_{ij} \nonumber \\
    &+ \sum_{n=1}^{N-1}\sum_{i}c_{i\alpha}(t_{n})c_{j\alpha}(t_{n}) \Xi(t_n)_{ij} \nonumber \\
    &+\frac{1}{2} \sum_{i} c_{i\alpha}(t_{N})c_{j\alpha}(t_{N}) \Xi(t_N)_{ij}
    ~,
\end{align}
where:
\begin{align}
  \Xi_{ij}(t_n) & = \lavg{\int_{t_n}^{t_{n+1}} \phi_{i}\phi_{j} dt} \nonumber \\
  & = \avg{ \phi_{i}\phi_{j}\Delta t}~
\end{align}
is a symmetric matrix as a function of time step.

The average of $\JIrr$ can be written as:
\begin{align}
    \lavg{{\JIrr}_{\alpha}} &=  \lavg{\int_{t_0}^{t_{1}} w \gamma dx_{\alpha}}+\cdots+\lavg{\int_{t_{N-1}}^{t_{N}} w \gamma dx_{\alpha}} \nonumber\\
    &-\frac{1}{2} \lavg{\int_{t_0}^{t_{1}}  dw_{\alpha}\cdot dp_{\alpha}}-\cdots-\frac{1}{2}\lavg{\int_{t_{N-1}}^{t_{N}} dw_{\alpha}\cdot dp_{\alpha}} \nonumber \\
    &=\sum_{n=0}^{N} \sum_{i} c_{i\alpha}(t_{n}) \mu_{i\alpha}(t_{n})
    ~,
\end{align}
where:
\begin{align}
    \mu_{i\alpha}(t_{n}) =& \langle\phi_{i}(x_{\alpha}(t_{n}),p_{\alpha}(t_{n}))\gamma dx_\alpha \\
    &- \frac{1}{2}\phi_{i}(x_{\alpha}(t_{n}),p_{\alpha}(t_{n}))(p_{\alpha}(t_{n+1})-p_{\alpha}(t_{n-1}))\rangle ,\\
    &~~~\text{for $n\neq0,N$};\\
    =& \langle\phi_{i}(x_{\alpha}(t_{0}),p_{\alpha}(t_{0}))\gamma dx_\alpha   \nonumber \\
    &+\frac{1}{2} \phi_{i}(x_{\alpha}(t_{0}),p_{\alpha}(t_{0}))(p_{\alpha}(t_{1})-p_{\alpha}(t_{0}))\rangle ,\\
    & ~~~\text{for $n=0$};~\text{and}\\
    =&-\avg{\frac{1}{2}\phi_{i}(x_{\alpha}(t_{N}),p_{\alpha}(t_{N}))(p_{\alpha}(t_{N})-P_{\alpha}(t_{N-1}))} , \\
    & ~~~\text{for $n=N$.}
\end{align}

Following the same algebra in the NESS state, the maximum of ratio $\mathcal{J}_{\alpha}$ is:
\begin{align}\label{eqn:nonNESSestresult}
   {\mathcal{J}_\alpha}_{\mathrm{max}}= \sum_{n=1}^{N}\sum_{i,j} \mu_{i\alpha}(t_{n})\big(\Xi(t_n)^{-1}\big)_{ij}\mu_{j\alpha}(t_{n})~.
\end{align}
This result Eq. \eqref{eqn:nonNESSestresult} clearly shows that even if we feed the long-time trajectories into this estimation scheme, the total EP estimated is equal to sum of EPs in each short duration $\Delta t$. This is why we call it \emph{rate-based estimation} when coefficients have time dependence. It is equivalent to estimating the entropy rate at each time step.

If our basis functions have time dependence:
\begin{align*}
    w_{\alpha}(\boldsymbol{x},\boldsymbol{p}) = \sum_{i}^{N} c_{i\alpha}\phi_{i}(\boldsymbol{x},\boldsymbol{p},t)~,
\end{align*}
the maximum ratio is similar to NESS state case. The variance of the stochastic current is:
\begin{align*}
\frac{1}{2D} \var{\JS} &= \lavg{\int_{\Gamma} w_\alpha^2 dt} \\
    &=\sum_{i,j} c_{i\alpha}c_{j\alpha} \lavg{\int_{\Gamma} \phi_{i}(t)\phi_{j}(t) dt} \\
    &=\sum_{i,j} c_{i\alpha}c_{j\alpha}\Xi_{ij} 
    ~,
\end{align*}
where:
\begin{align*}
    \Xi_{ij}=\lavg{\int_{\Gamma} \phi_{i}(t)\phi_{j}(t) dt}~.
\end{align*}
And, the discretized version is:
\begin{align*}
    \Xi_{ij}=\lavg{\sum_{n=0}^{N-1} \phi_{i}(\boldsymbol{x}(t_{n}), \boldsymbol{p}(t_{n}),t_{n})\phi_{j}(\boldsymbol{x}(t_{n}), \boldsymbol{p}(t_{n}),t_{n}) \Delta t}
\end{align*}
in starting point discretization or:
\begin{align*}
    \Xi_{ij}=\lavg{\sum_{n=1}^{N} \phi_{i}(\boldsymbol{x}(t_{n}), \boldsymbol{p}(t_{n}),t_{n})\phi_{j}(\boldsymbol{x}(t_{n}), \boldsymbol{p}(t_{n}),t_{n}) \Delta t}
\end{align*}
in ending-point discretization.

The average of the irreversible current in $\alpha-$th direction is:
\begin{align*}
    \avg{{\JIrr}_{\alpha}} &= \lavg{\int_{\Gamma} w \gamma dx_\alpha-\frac{1}{2} \int_{\Gamma} dw_\alpha \cdot dp_\alpha} \\
    &=\sum_{i} c_{i\alpha} \lavg{\int_{\Gamma} \phi_{i} \gamma dx -\frac{1}{2} \int_{\Gamma} d\phi_{i} \cdot dp_\alpha} \\
    &=\sum_{i} c_{i\alpha}\mu_{i\alpha}~,
\end{align*}
where:
\begin{align}
    \mu_{i\alpha} = \lavg{\int_{\Gamma} \phi_{i} \gamma dx -\frac{1}{2} \int_{\Gamma} d\phi_{i} \cdot dp_\alpha}~.
\end{align} 
The discretized version is:
\begin{align*}
    \mu_{i\alpha} = & \left\langle \sum_{n=0}^{N-1} \phi_{i}(\boldsymbol{x}(t_{n}),\boldsymbol{p}(t_{n}),t_{n}) \gamma (x_{\alpha}(t_{n+1})-x_{\alpha}(t_{n}))-\nonumber \right. \\
    &\frac{1}{2} (\phi_{i}(\boldsymbol{x}(t_{n+1}),\boldsymbol{p}(t_{n+1}),t_{n+1})-\phi_{i}(\boldsymbol{x}(t_{n}),\boldsymbol{p}(t_{n}),t_{n})) \nonumber \\
    & \times (p_\alpha (t_{n+1}) - p_\alpha (t_{n})) \big\rangle ~.
\end{align*}
The maximum of $\mathcal{J}_{\alpha}$ is then:
\begin{align}
    {\mathcal{J}_\alpha}_{\mathrm{max}} = \sum_{i,j} \mu_{i\alpha} \Xi^{-1}_{ij} \mu_{j\alpha}~.
\end{align}

Those above are all for EP in one direction. To account the total EP, we need to sum over all directions:
\begin{align}
    \Sigma = \sum_{\alpha} {\mathcal{J}_\alpha}_{\mathrm{max}}~.
\end{align}

For all scenarios mentioned above, we encounter the problem finding the maximum of a ratio in the form of:
\begin{align}
    \mathcal{J} = \frac{\sum_{i,j}c_{i}c_{j}\mu_{i}\mu_{j}}{\sum_{i,j}c_{i}c_{j}\Xi_{ij}} ~
\end{align}
with respect to $c_{i}$. Since this ratio is invariant under rescaling $c_{i}$: $c_{i}\to \kappa c_{i}$, we can set $\sum_{i} c_{i}\mu_{i}=1$. The challenge turns out to be finding the minimum of $\sum_{i,j}c_{i}c_{j}\Xi_{ij}$ given the constraint $\sum_{i} c_{i}\mu_{i}=1$. To avoid overfitting when the number of estimators---or the dimension of $c_{i}$---is large compared to the size of the data set, we include regularization. The Lagrangian function now is:
\begin{align*}
    \mathcal{L} &=  \sum_{i,j}c_{i}c_{j}\Xi_{ij} + \beta \sum_{i} c_{i}^2 + \lambda (\sum_{i} c_{i}\mu_{i}-1)\\
    &=\sum_{i,j}c_{i}c_{j}(\Xi_{ij} +\beta \delta_{ij}) + \lambda (\sum_{i} c_{i}\mu_{i}-1)~.
\end{align*}
With L-2 regularization, the new result is:
\begin{align}
     \sum_{i,j} \mu_{i} \Xi^{-1}_{ij} \mu_{j} \to \sum_{i,j} \mu_{i} (\Xi+\beta I)^{-1}_{ij} \mu_{j}~.
\end{align}
We can expand this result in powers of $\beta$: 
\begin{align}
    \sum_{i,j} \mu_{i} (\Xi+\beta I)^{-1}_{ij} \mu_{j} &\sim \sum_{i,j} \mu_{i} \Xi^{-1}_{ij} \mu_{j}- \beta \sum_{i,j} \mu_{i} \Xi^{-2}_{ij} \mu_{j}\nonumber  \\
    &+ \beta^2\sum_{i,j} \mu_{i} \Xi^{-3}_{ij} \mu_{j} + \mathcal{O}(\beta^3)~.
\end{align}
Assuming $\beta$ is small, the regularization decrease the unregularized result. This bias was minor for all cases tested---changing the relative error by no more than half a percent. See Figs. \ref{fig:rate_unreg} and \ref{fig:cyclic_NESS_unreg} which compare unregularized and regularized data sets. For all cases, we set $\beta = N^{-2}$.

\section{Model details}
\label{appendix:freediffusiondetails}
\subsection{Free diffusion}

This appendix presents the main results for free diffusion EP in the underdamped regime. The Langevin equations are:
\begin{align}\label{eqn:freediffusioneom}
    & dx = p/m ~dt~,~ \\
    & dp = - \gamma p/m ~dt + dB_{t}~.
\end{align}
And, the probability distribution $f(x,p,t)$ evolution obeys the Krammer's equation:
\begin{align}\label{eqn:Krammerfreediffusion}
    \partial_{t}f + \partial_{x}(p/m f) + \partial_{p}(-\gamma p/m - \gamma T \partial_{p}f/f) = 0~.
\end{align}
The Green function of Eq. \eqref{eqn:Krammerfreediffusion} is:
\begin{align}
    f(x,p,x',p',t)=&{\frac {1}{2\pi \sigma _{X}\sigma _{P}{\sqrt {1-\beta ^{2}}}}} \exp \Bigg(-{\frac {1}{2(1-\beta ^{2})}}\nonumber \\
    &[{\frac {(x-\mu _{X})^{2}}{\sigma _{X}^{2}}}+{\frac {(p-\mu _{P})^{2}}{\sigma _{P}^{2}}}\nonumber \\
    &-{\frac {2\beta (x-\mu _{X})(p-\mu _{P})}{\sigma _{X}\sigma _{P}}}] \Bigg),
\end{align}
where:
\begin{align*}
\sigma_{X}^{2} & ={\frac {mT}{\gamma ^{2}}}\left[1+2\frac{\gamma}{m} t-\left(2-e^{-\frac{\gamma}{m} t}\right)^{2}\right]\\ 
\sigma _{P}^{2} & =m T\left(1-e^{-2\frac{\gamma}{m}t}\right)\\
\beta  & ={\frac {mT}{{\gamma} \sigma_{X}\sigma _{P}}}\left(1-e^{-\frac{\gamma}{m} t}\right)^{2}\\
\mu _{X} & =x'+\frac{1}{\gamma} \left(1-e^{-\frac{\gamma}{m} t}\right)p' ~,~\text{and}\\ 
\mu _{P} & =p'e^{-\frac{\gamma}{m} t}~.
\end{align*}
If the initial probability distribution is a Gaussian distribution that is uncorrelated in position and momentum, i.e., 
\begin{align}
    f(x,p,0) = \frac{1}{2 \pi \sigma_{x_0}\sigma_{p_0}} e^{-\frac{(x-x_{0})^2}{2\sigma_{x_0}^2}-\frac{(p-p_{0})^2}{2\sigma_{p_0}^2}}
    ~,
\end{align}
then the distribution is always Gaussian with mean $\boldsymbol{\mu}$ and variance $\boldsymbol{\Sigma}$, where:
\begin{align*}
    \boldsymbol{\mu}(t) &= (\mu_{x},\mu_{p})^{\top}  \\
    \boldsymbol{\Sigma}(t) &= 
    \begin{pmatrix}
        \sigma_x^2 & \mathrm{cov}(x,p) \\
        \mathrm{cov}(x,p) & 
        \sigma_p^2
    \end{pmatrix}~
\end{align*}
with:
\begin{align*}
    \mu_{x}&=x_{0} + \frac{1}{\gamma}(1-e^{-\frac{\gamma}{m}t})p_{0} \\
    \mu_{p}&=p_{0}e^{-\frac{\gamma}{m}t}~
\end{align*}
and:
\begin{align*}
    \mathrm{cov}(x,p)&=\frac{1}{\gamma}  (1  -e^{-\frac{\gamma}{m}t}) ( m T -m Te^{- \frac{\gamma}{m}t}  + \sigma_{p_0}^2 e^{- \frac{\gamma}{m}t}) \\
    \sigma_p^2 &= (\sigma_{p_0}^2 - mT) e^{-2\frac{\gamma}{m}t} +mT \\
    \sigma_x^2 &= -\frac{1}{\gamma^2} \left(1-e^{-\frac{\gamma }{m}t}\right)\nonumber \\
    & \quad\times  \left(mT
   (3 -e^{-\frac{\gamma}{m}t})-\sigma_{p_0}^2 (1-e^{-\frac{\gamma}{m}t})\right)\nonumber\\
   &\quad+\frac{2 T }{\gamma }t+\sigma_{x_0}^2~.
\end{align*}
To compute the EP, the heat $Q$ transferred to the bath is equal to the negative change in kinetic energy:
\begin{align*}
    Q = -\frac{1}{2m}(\mu_{p}^2 + \sigma_p^2-p_{0}^2-\sigma_{p_{0}}^2)~.
\end{align*}
The EP rate is given by:
\begin{align*}
    \Sigma &= \partial_{t} \left(Q/T + \frac{1}{2 }\log \det{\boldsymbol{\Sigma}}(t)-\frac{1}{2 }\log \det{\boldsymbol{\Sigma}}(0)\right)\nonumber \\
           &= \partial_{t}\left(-\mu_{p}^2 - \sigma_p^2 + \frac{1}{2 }\log \det{\boldsymbol{\Sigma}}\right)~.
\end{align*}
The $\uIrr$ in this case is:
\begin{align*}
    \uIrr &= -\gamma\frac{p}{m} - \gamma T \partial_{p}\log f(x,p,t)\nonumber\\
    &= \left( -\frac{\gamma}{m} - \frac{\gamma T}{(-1+\beta^2)\sigma_{p}^2} \right) p \nonumber \\
    &\quad+\frac{\gamma T \beta}{(-1+\beta^2)\sigma_{x}\sigma_{p}} x +\frac{\gamma T \mu_{p}}{(-1+\beta^2)\sigma_{p}^2} \nonumber \\
    &\quad-\frac{\gamma T \beta \mu_{x}}{(-1+\beta^2)\sigma_{x}\sigma_{p}}~.
\end{align*}

\subsection{Particle on a ring}
\label{app:ParticleOnRing}
\begin{figure*}[t]
    \begin{subfigure}{1\textwidth}
        \centering
        \caption{}
        \scalebox{.85}{
        \includegraphics{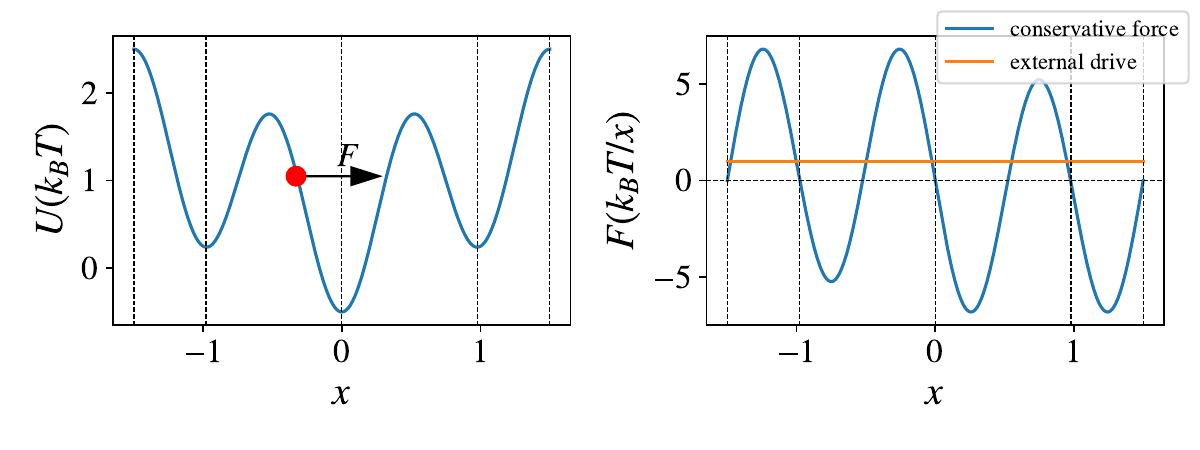}
        }
        \label{fig:cyclicsketch}
    \end{subfigure} \\
    \begin{subfigure}{1\columnwidth}
        \centering
        \caption{}
        \scalebox{0.55}{
        \includegraphics{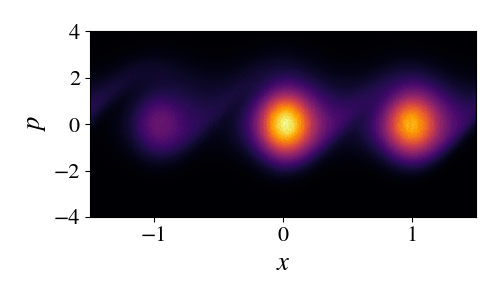}
        }
        \label{fig:cyclicpdf}
    \end{subfigure}
    \caption{(a) Sketch of the NESS model: An underdamped particle trapped in a ring of length $3$ driven by a constant force $F$ within a cyclic potential: $V(x+3)=V(x)$. (b) NESS probability distribution of the particle.} 
    \label{fig:cyclic_NESS}
\end{figure*}

In this model, we study a particle on a ring of circumference 3 driven by a constant force of $F=1$ (in units of $k_B T$ per distance). The potential energy around the ring is composed of three sinusoidal wells. Two of these wells are shallow on one side, and one of them is deep on both sides.  With sufficiently long time, the system reaches a NESS state. Figures \ref{fig:cyclicsketch} and \ref{fig:cyclicpdf} show the potential and NESS probability distribution in $(x,p)$ phase space. The depths of the shallow and deep barriers are $~1.5 k_B T$ and $~2.25 k_B T$, respectively.

\begin{figure*}[!t]
    \centering
    \includegraphics[width=\textwidth]{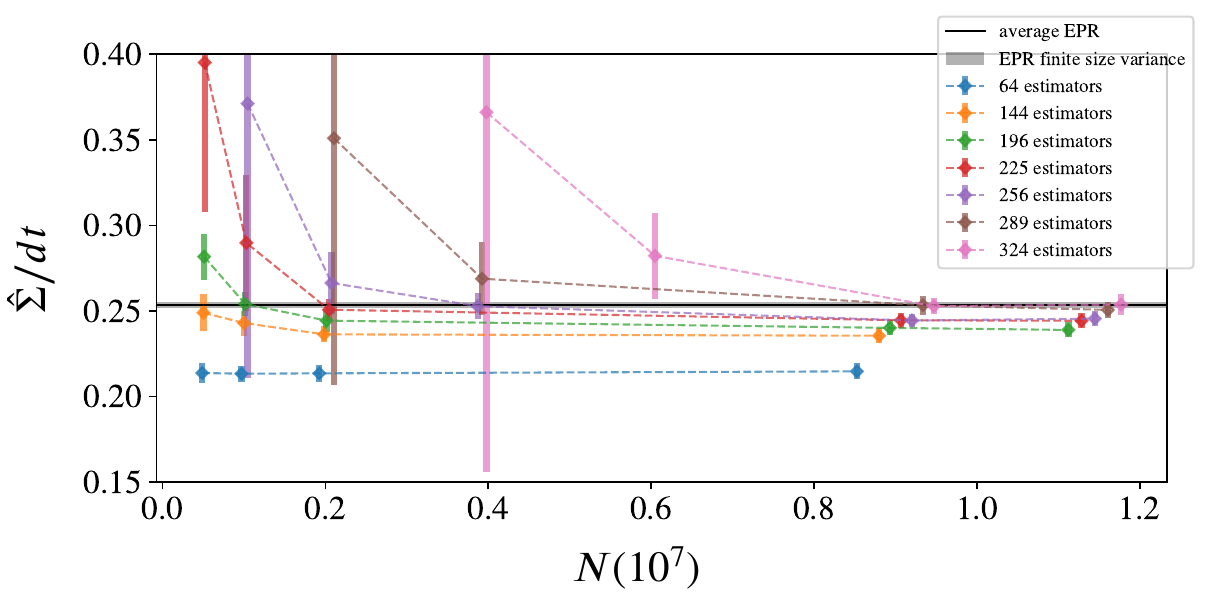}
    \caption{Unregularized entropy production rate estimation for the cyclic NESS. Here, we must go to $~10^7$ trajectories to converge to the correct value when the data is not regularized. The key parts of the algorithm scale with $(\text{num estimators})^2(\text{num paths})$ and $(\text{num estimators})^6$. The first term tends to dominate because so much data is needed and the computation becomes computationally quite expensive.}
    \label{fig:cyclic_NESS_unreg}
\end{figure*}

\section{Estimator details}
\label{appendix:detailsonestimators}

This appendix discusses how to determine parameters in Gaussian kernels. We first consider kernels for the rate-base estimation at time $t$. For an $m$-dimensional system, the phase space dimension is $2m$. We need to approximate an $m$-dimensional weight $\boldsymbol{w}(\boldsymbol{x}, \boldsymbol{p}, t )$. We denote its $\alpha$ component as $w_{\alpha}(\boldsymbol{x}, \boldsymbol{p}, t)$. For each component, we use the same $k$ Gaussian kernels to approximate it; denoted as $G_{i}(\boldsymbol{x},\boldsymbol{p})$. The Gaussian kernels have the form: 
\begin{align*}
    G_{i}(\boldsymbol{x},\boldsymbol{p}) = e^{-(\boldsymbol{x}-\boldsymbol{\bar{x}}_{i})^{\top}\Sigma_{x}^{-1}(\boldsymbol{x}-\boldsymbol{\bar{x}}_{i})-(\boldsymbol{p}-\boldsymbol{\bar{p}}_{i})^{\top}\Sigma_{p}^{-1}(\boldsymbol{p}-\boldsymbol{\bar{p}}_{i})}
    ~,
\end{align*}
where $(\boldsymbol{\bar{x}}_{i}^{\top},\boldsymbol{\bar{p}}_{i}^{\top}$) are the $i$-th kernel's center in the phase space. $\Sigma_{x}$ and $\Sigma_{p}$ are the bandwidth of Gaussian kernels---two diagonal positive diagonal matrices. We choose centers to be on a rectangular grid. Along each grid direction, we set $n$ points and $k=n^{2m}$.

We first determine the maximum and minimum in phase space of all trajectories as  $\boldsymbol{x}_{\min}$, $\boldsymbol{x}_{\max}$, $\boldsymbol{p}_{\min}$, and $\boldsymbol{p}_{\max}$. For each direction $q$ in phase space $(q=x_{\alpha},p_{\alpha})$, the bandwidth is determined by $\Sigma_{q} = (q_{\max}-q_{\min}+\sigma_q)/(n-2)$, where $\sigma_q$ is the standard deviation of trajectories data of $q$ at time $t$. And, the centers in the $q$ direction are $\bar{q}_\alpha =\{ q_{\min} -\sigma_{q} + (j-\frac{1}{2})\cdot \sigma_{q}\}_{j=0}^{n}$. The centers lie on a $2m$-dimension grid of each direction's centers.

For one-shot estimation, we add time dependence to the Gaussian kernels:
\begin{align*}
    G_{i}(\boldsymbol{x},\boldsymbol{p},t) =& e^{-\frac{1}{2}(\boldsymbol{x}-\boldsymbol{\bar{x}}_{i})^{\top}\Sigma_{x}^{-1}(\boldsymbol{x}-\boldsymbol{\bar{x}}_{i})-\frac{1}{2}(\boldsymbol{p}-\boldsymbol{\bar{p}}_{i})^{\top}\Sigma_{p}^{-1}(\boldsymbol{p}-\boldsymbol{\bar{p}}_{i})}\nonumber \\
    &\times e^{-(t-t_{i})^2/2\sigma_{t}^2} ~.
\end{align*}
In the time direction, the number of estimators is chosen to be $n_{t}$. We let our estimators cover a wider range than $[0,\tau]$. The bandwidth and centers are set to be $\sigma_{t} = \tau/ (n_{t}-2)$ and $\{-\frac{1}{2}\sigma_{t}+(i-1)\sigma_{t}\}_{i=1}^{n_{t}}$, respectively. 

The number $n$ of estimators in each direction can be tuned in each direction. If $n$ is too small, the Gaussian kernels cannot capture the behavior of the real weights and the estimates converge to a quantity less than the actual EP. Large $n$, however, leads to overfitting and this requires more trajectories to converge. For each different case, we used different numbers of estimators.

Figure \ref{fig:cyclic_NESS_unreg} shows how different numbers of estimators converge for different numbers of paths for the particle on a ring NESS system. This plot is using an un-regularized method, to provide comparison with Fig. \ref{fig:cyclic_NESS18N126} in the main text. We see that the general trend is the same as Fig. \ref{fig:cyclic_NESS18N126}, but convergence is messier and slower. A general trend is that this estimation scheme underestimates the EP (rate) when the number of estimators is too small. With a finite number of trajectories, however, a large high number of estimators encounters overfitting. For a fixed number of estimators, we  see that the standard error decreases as number of trajectories increases. With more estimators, the average estimation is closer to the real value of EP rate.

\section{Simulation Detail}\label{appendix:simulation}

The free diffusion simulation used the Euler-Maryama integration method to sample trajectories. The cyclic NESS simulations additionally used 4th-order Runge-Kutta for the deterministic portion of the integration. For the cyclic NESS system, the steady state was approximated by running the simulation until stationarity was reached. For all simulations, Python's NumPy random generator was used to produce realizations of $dB_t$. In all estimation problems, $dt=10^{-3}$, with $\gamma=k=m=T=1$. For the NESS system, to assure faithful dynamics, trajectories were simulated using $dt=2\times10^{-4}$---but only every $5^\text{th} $ step was kept to simulate trajectories measured with a time interval of $10^{-3}$.

\end{document}